\documentclass[12pt]{article}
\usepackage{amsmath}  

\begin{document}
\renewcommand{\theequation}{\thesection.\arabic{equation}}
\def\prg#1{\medskip{\bf #1}}     \def\ra{\rightarrow}
\def\lra{\leftrightarrow}        \def\Ra{\Rightarrow}
\def\nin{\noindent}              \def\pd{\partial}
\def\dis{\displaystyle}          \def\inn{\,\rfloor\,}
\def\grl{{GR$_\Lambda$}}         \def\vsm{\vspace{-9pt}}
\def\Lra{{\Leftrightarrow}}      \def\ads3{AdS$_3$}
\def\cs{{\scriptscriptstyle \rm CS}}  \def\ads3{{\rm AdS$_3$}}
\def\Leff{\hbox{$\mit\L_{\hspace{.6pt}\rm eff}\,$}}
\def\bull{\raise.25ex\hbox{\vrule height.8ex width.8ex}}
\def\ric{{(Ric)}}         \def\tmgl{\hbox{TMG$_\Lambda$}}
\def\km{$u(1)_{\rm KM}$}  \def\tmg{{\rm TMG}}
\def\vkm{Virasoro$\,\oplus_{\rm sd}\,u(1)_{\rm KM}$}
\def\sdplus{\oplus_{sd}}
\def\cs{{\rm CS}}     \def\gr{{\rm G}}      \def\reg{{\rm r}}

\def\G{\Gamma}        \def\S{\Sigma}        \def\L{{\mit\Lambda}}
\def\a{\alpha}        \def\b{\beta}         \def\g{\gamma}
\def\d{\delta}        \def\m{\mu}           \def\n{\nu}
\def\th{\theta}       \def\k{\kappa}        \def\l{\lambda}
\def\vphi{\varphi}    \def\ve{\varepsilon}  \def\p{\pi}
\def\r{\rho}          \def\Om{\Omega}       \def\om{\omega}
\def\s{\sigma}        \def\t{\tau}          \def\eps{\epsilon}
\def\nab{\nabla}

\def\bA{{\bar A}}     \def\bF{{\bar F}}     \def\bG{{\bar G}}
\def\bt{{\bar\tau}}   \def\bg{{\bar g}}     \def\tG{{\tilde G}}
\def\cL{{\cal L}}     \def\cM{{\cal M }}    \def\cE{{\cal E}}
\def\cH{{\cal H}}     \def\bcH{{\bar\cH}}   \def\hcH{\hat{\cH}}
\def\cK{{\cal K}}     \def\hcK{\hat{\cK}}
\def\cO{{\cal O}}     \def\hcO{\hat{\cal O}}
\def\cB{{\cal B}}     \def\heps{\hat\epsilon}
\def\bu{{\bar u}}     \def\bw{{\bar w}}     \def\bk{{\bar\k}}
\def\bpi{{\bar\pi}}   \def\bphi{{\bar\phi}} \def\tI{{\tilde I}}

\def\hA {{\hat A}}      \def\hB{{\hat B}}   \def\hu{{\hat u}}
\def\hOm{{\hat\Omega}}  \def\hL{{\hat\L}}   \def\hxi{{\hat\xi}}
\def\hhB{{\hat\hB}{}}   \def\hhL{{\hat\hL}{}}  \def\hth{{\hat\th}}
\def\hhOm{{\hat\hOm}{}} \def\hd{{\hat\delta}}  \def\hhxi{{\hat\hxi}}
\def\nn{\nonumber}
\def\be{\begin{equation}}             \def\ee{\end{equation}}
\def\ba#1{\begin{array}{#1}}          \def\ea{\end{array}}
\def\bea{\begin{eqnarray} }           \def\eea{\end{eqnarray} }
\def\beann{\begin{eqnarray*} }        \def\eeann{\end{eqnarray*} }
\def\beal{\begin{eqalign}}            \def\eeal{\end{eqalign}}
\def\lab#1{\label{eq:#1}}             \def\eq#1{(\ref{eq:#1})}
\def\bsubeq{\begin{subequations}}     \def\esubeq{\end{subequations}}
\def\bitem{\begin{itemize}}           \def\eitem{\end{itemize}}

\title{Asymptotic Chern-Simons formulation\\
       of spacelike stretched AdS gravity}

\author{M. Blagojevi\'c and B. Cvetkovi\'c\footnote{
        Email addresses: {\tt mb@ipb.ac.rs,
                                cbranislav@ipb.ac.rs}} \\
University of Belgrade, Institute of Physics,\\ P. O. Box 57, 11001
Belgrade, Serbia}
\date{}
\maketitle
\begin{abstract}
We show that the asymptotic structure of topologically massive gravity
in the spacelike stretched AdS sector can be faithfully represented by
an $SL(2,R)\times U(1)$ Chern-Simons gauge theory, by adopting a
natural correspondence between their fields and coupling constants.
\end{abstract}

\section{Introduction} 

In three-dimensional (3D) Einstein's gravity with a cosmological
constant (\grl), the AdS asymptotic structure is described by two
independent Virasoro algebras with classical central charges
\cite{1,2}. Of particular importance for our understanding of the
corresponding gravitational dynamics, at both classical and quantum
level, is the fact that \grl\ can be represented as an ordinary gauge
theory---the Chern-Simons (CS) theory based on the internal
$SL(2,R)\times SL(2,R)$ gauge group \cite{3,4}.

Adding the gravitational CS term to \grl\ substantially changes its
dynamical structure: while \grl\ is a topological theory with no
dynamical degrees of freedom, the new theory, known as topologically
massive gravity with a cosmological constant (\tmgl), is a truly
dynamical theory with one degree of freedom, the massive graviton
\cite{5}. In \grl, the AdS sector is defined around the maximally
symmetric vacuum \ads3, and it contains the BTZ black hole with
interesting thermodynamic properties \cite{6,7}. Since the same \ads3\
is authomatically a solution of \tmgl, one can define the AdS sector
also in \tmgl. However, for generic values of the coupling constants,
the physical interpretation of this sector suffers from serious
difficulties: for the usual sign of the gravitational coupling constant
$G>0$, massive excitations around \ads3\ carry negative energies
\cite{5}, while for $G<0$, the black hole mass becomes negative
\cite{8,9}.

In an interesting attempt to find a resolution of this
inconsistency, Li et al \cite{9} studied the chiral version of
\tmgl, defined by a specific relation between the graviton mass
and the cosmological constant. However, we shall focus our
attention to another promising idea: Anninos et al \cite{10}
suggested that choosing a new vacuum, the spacelike stretched
\ads3, could lead to a stable ground state of \tmgl.
Geometrically, choosing the spacelike stretched \ads3\ as the
ground state corresponds to a deformation of the \ads3\ isometry
group $SL(2,R)\times SL(2,R)$ to its four-parameter subgroup
$SL(2,R)\times U(1)$ \cite{11,12}.

The constrained Hamiltonian analysis of the full \tmgl\ was carried out
recently in \cite{13}, see also \cite{14}, leading to a clear and
precise picture of its gauge and dynamical features. An important step
towards a proper understanding of the spacelike stretched AdS
asymptotic structure was achieved by constructing a set of suitable
asymptotic conditions \cite{15,16}. The resulting asymptotic symmetry
was shown to be centrally extended semidirect sum of a Virasoro and a
$u(1)$ Kac-Moody algebra, \vkm. This whole sector of \tmgl\ will be
called shortly the spacelike stretched AdS gravity. Motivated by the
experience stemming from \grl, the goal of the present paper is to
improve our understanding of the spacelike stretched AdS gravity by
showing that its {\it asymptotic\/} structure can be faithfully
represented by an $SL(2,R)\times U(1)$ CS gauge theory.

Let us mention here a highly interesting hypothesis formulated by
Anninos et al \cite{10}, according to which the spacelike stretched AdS
boundary dynamics is characterized by a two-dimensional conformal
symmetry with central charges. The proof of this hypothesis, presented
recently in \cite{16}, is based on the asymptotic \vkm\ canonical
algebra and the (algebraic) Sugawara construction, see also \cite{17}.
We expect that the asymptotic CS representation of the spacelike
stretched AdS gravity will be a useful tool in clarifying the boundary
conformal structure lying behind the Sugawara construction.

The paper is organized as follows. In Sections 2 and 3, we use the
canonical approach to study the asymptotic structure of the
$SL(2,R)\times U(1)$ CS gauge theory, a natural counterpart of the
spacelike stretched AdS gravity. The result of this analysis is the
\vkm\ Poisson bracket algebra of the canonical generators. Then, in
Section 4, we give a brief overview of the basic asymptotic features of
the spacelike stretched AdS gravity, including its asymptotic canonical
algebra, \vkm. In section 5 we introduce specific asymptotic
conditions,   find a new form of the canonical surface terms and count
and identify boundary degrees of freedom of the spacelike stretched AdS
gravity. In Section 6, we compare the resulting gravitational
asymptotic structure with the one found in the CS theory and find, by a
natural identification of the corresponding coupling constants and
dynamical fields, that they are identical. Section 7 is devoted to
concluding remarks, while appendices contain some technical details.

Our conventions are the same as in Ref. \cite{16}: the Latin indices
refer to both the basis of $sl(2,R)$ and the local Lorentz frame, the
Greek indices refer to the coordinate frame;  the middle alphabet
letters $(i,j,k,...;\m,\n,\l,...)$ run over 0,1,2, the first letters of
the Greek alphabet $(\a,\b,\g,...)$ run over 1,2; the metric components
in the local Lorentz frame are $\eta_{ij}=(+,-,-)$; totally
antisymmetric tensor $\ve^{ijk}$ and the related tensor density
$\ve^{\m\n\r}$ are both normalized as $\ve^{012}=1$.

\section{{\boldmath $SL(2,R)\times U(1)$} Chern-Simons gauge theory}
\setcounter{equation}{0} 

Motivated by the asymptotic structure of \tmgl\ in the spacelike
stretched AdS sector \cite{16}, we discuss here the corresponding
aspects of the $SL(2,R)\times U(1)$ CS gauge theory.

\subsection{The action and boundary conditions}

Consider the CS gauge theory defined by the action
\be
I_\cs=-\k\int_{\cM}\Bigl(A^idA_i+\frac{1}{3}\ve_{ijk}A^iA^jA^k\Bigr)
      +\bk\int_{\cM}\bA d\bA\, .                           \lab{2.1}
\ee
Here, $\cM$ is a spacetime manifold with topology $R\times \S$, where
$R$ is interpreted as time and $\S$ is a spatial manifold whose
boundary is topologically a circle (which may be located at infinity),
$A^i=A^i{_\m}dx^\m$ and  $\bA=\bA_\m dx^\m$ (1-forms) are the $SL(2,R)$
and $U(1)$ gauge potentials, respectively, and $\ve_{ij}{^k}$ are the
structure constants of $sl(2,R)$ (Appendix A). The action is invariant
under the infinitesimal gauge transformations
$$
\d_0 A^i=\nabla u^i:=du^i+\ve^i{}_{jk}A^j u^k\, ,\qquad
\d_0\bA=d\bu\, ,
$$
where $u^i$ and $\bu$ are gauge parameters.

We assume the existence of the Schwarzschild-like coordinates
$x^\m=(t,\r,\vphi)$ on $\cM$, such that the boundary $\pd\S$ is
described by the standard angular coordinate $\vphi$. Having in mind
the fact that the asymptotic parameters in the spacelike stretched AdS
sector of \tmgl\ are \emph{time independent}, we choose the CS boundary
conditions as
\be
A^i{}_0=0\, ,\qquad\bA_0=\bar a_0\qquad \mbox{at~~}\pd\S\,,\lab{2.2}
\ee
since they imply $\pd_0 u^i=0$, $\pd_0\bu=0$ at the boundary. Although
the choice \eq{2.2} leads to a nontrivial boundary term $\d B$ in the
variation of the action, this can be corrected by introducing the
improved action $\tilde I:= I-B$, which produces the standard field
equations:
\be
F^i:=dA^i+\ve^i{}_{jk}A^jA^k=0\, ,\qquad \bF:=d\bA=0\, .
\ee
Consequently, the Lie algebra valued gauge potentials are locally
trivial: $A_\m=G^{-1}\pd_\m G$, $\bA_\m=\bG^{-1}\pd_\m\bG$, where $G$
and $\bG$ are elements of $SL(2,R)$ and $U(1)$, respectively.

\subsection{The canonical structure}

Now, we analyze the symmetry structure of our CS theory by using the
canonical formalism.

\prg{Gauge generator.} Introducing the canonical momenta
$(\pi_i{^\m},\bpi^\m)$ corresponding to the Lagrangian variables
$(A^i{_\m},\bA_\m)$, one obtains the primary constraints:
\bea
&&\pi_i{^0}\approx 0\,,\qquad
  \phi_i{^\a}:=\pi_i{^\a}+\k\ve^{0\a\b}A_{i\b}\approx 0\, ,\nn\\
&&\bpi^0\approx 0\,,\qquad
  \bphi^\a:=\bpi^\a-\bk\ve^{0\a\b}\bA_\b\approx 0\, ,      \nn
\eea
where  $\a,\b=1,2$. The secondary constraints have the form
\bea
&&\cH_i:=\k\ve^{0\a\b}F_{i\a\b}-\nabla_\a\phi_i{^\a}\approx 0\,,\nn\\
&&\bcH:=-\bk\ve^{0\a\b}\bF_{\a\b}-\pd_\a\bphi{^\a}\approx 0\,,\nn
\eea
and the total Hamiltonian (up to an irrelevant divergence) is given by
\be
\cH_T=A^i{_0}\cH_i+\bA^0\bcH+w^i{_0}\pi_i{^0}+\bw\bpi^0\,, \nn
\ee
where $w^i$ and $\bw$ are arbitrary multipliers.

The constraints $(\pi_i{^0},\cH_i,\bpi^0,\bcH)$ are first class while
$(\phi_i{^\a},\bphi^\a)$ are second class. Using Castelanni's procedure
\cite{18}, one finds the form of the canonical gauge generator:
\be
G=\int d^2 x \left[(\nabla_0u^i)\pi_i{^0}+u^i\cH_i\right]
  +\int d^2 x \left[(\pd_0\bu)\bpi^0+\bu\bcH\right]\, .    \lab{2.4}
\ee

\prg{Fixing the gauge.} We have found two sets of the first class
constraints, $(\pi_i{^0},\bpi^0)$ and $(\cH_i,\bcH)$. The first set of
the corresponding gauge conditions is chosen so as to extend the
boundary conditions \eq{2.2} to the whole spacetime:
\bsubeq\lab{2.5}
\be
A^i{}_0=0\,,\qquad \bA_0=\bar a_0\, .                        \lab{2.5a}
\ee
Using the notation $A_\m=A^i{_\m}T_i$, where $T_i$ is a basis of the
$sl(2,R)$ Lie algebra (Appendix A), the second set of gauge conditions
is defined by restricting $A_1$ and $\bA_1$ to be functions of the
radial coordinate only:
$$
A_1\approx b^{-1}(\r)\pd_1 b(\r)\, ,\qquad
\bA_1\approx\bar b^{-1}(\r)\pd_1 \bar b(\r)\, ,
$$
where $b$ and $\bar b$ are in $SL(2,R)$ and $U(1)$, respectively.
By a suitable choice of the radial coordinate, we can write
\be
A_1=a_1\, ,\quad \bA_1=\bar a_1\, ,                         \lab{2.5b}
\ee
\esubeq
where $a_1=a^i_1T_i$ and $\bar a_1$ are constant elements of the
corresponding Lie algebras. The gauge conditions \eq{2.5b} are
conserved in time.

The constraints $F_{12}\approx 0$, $\bF_{12}\approx 0$ imply:
\be
A_2\approx b^{-1}\hA_2(t,\vphi)b\, ,\qquad \bA_2\approx\bA_2(t,\vphi)\, ,\nn
\ee
while the field equations $F_{02}=0\,,\bF_{02}=0$ lead to:
$$
\hA_2=\hA_2(\vphi)\,,\qquad \bA_2=\bA_2(\vphi)\,.
$$
The residual gauge symmetry has the form:
\be
\d_0\hA^i{}_2=\pd_2\hu^i+\ve^i{}_{jk}\hA^j{_2}\hu^k\, ,\qquad
\d_0\bA_2=\pd_2\bu\, ,                                     \lab{2.6}
\ee
where $u=:b^{-1}\hu(\vphi)b$ and $\bu=\bu(\vphi)$.

\prg{The improved generator.} After adopting the gauge conditions
\eq{2.5}, the effective gauge generator can be written as
$$
G=\int d^2 x u^i\cH_i+\int d^2 x \bu\bcH \, .
$$
This expression is not differentiable. Indeed, when the gauge
parameters are independent of field derivatives, the variation of $G$
contains certain boundary contributions:
\bsubeq
\bea
&&\d G=-\d\G_L[u]-\d\G_R[\bu]+R\, ,                        \nn\\
&&\d\G_L[u]=\oint df_\a u^i\left(
  -2\k\ve^{0\a\b}\d A_{i\b}+\d\phi_i{^\a}\right)\, ,       \nn\\
&&\d\G_R[\bu]=\oint df_\a \bu\left(
   2\bk\ve^{0\a\b}\d\bA_\b+\d\bar\phi^\a\right)\, .        \lab{2.7a}
\eea
Here, $R$ are regular terms, which correspond to well-defined
functional derivatives, $\d\G_L,\d\G_R$ are the boundary integrals
obtained with the help of the Stokes theorem, and
$df_\a=\ve_{0\a\b}dx^\b$ represents the line element on the spatial
boundary. If we can integrate $\d\G_L[u]$ and $\d\G_R[\bu]$ to find
$\G_L[u]$ and $\G_R[\bu]$, the improved gauge generator takes the form
\be
\tG\left[u,\bu\right]=G\left[u,\bu\right]+\G_L[u]+\G_R[\bu]\,.\lab{2.7b}
\ee
\esubeq

\section{Asymptotic symmetries}
\setcounter{equation}{0} 

The form of the improved canonical generator $\tG$, which gives a
complete description of the boundary symmetry, depends on the boundary
conditions imposed on $u^i$ and $\bu$.

\subsection{Kac-Moody extension of {\boldmath $sl(2,R)\oplus u(1)$}}

The simplest boundary conditions on the gauge parameters that allows us
to find an explicit form of the surface terms is defined as follows:
\bitem
\item[\bull] $u$ and $\bu$ are independent of the fields
         (at the boundary).
\eitem
On the subspace defined by $\phi_i{^\a},\bar\phi^\a\approx 0$, the
relations \eq{2.7a} imply
\bea
&&\G_L[\t]\approx -2\k\oint dx^\b \t^i A_{i\b}\, ,         \nn\\[3pt]
&&\G_R[\bar\t]\approx 2\bk\oint dx^\b \bar\t\bA_\b\, .     \lab{3.1b}
\eea

To find the PB algebra of the improved generator \eq{2.7b}, we use the
relation
$$
\{\tG[\t,\bar\t],\tG[\l,\bar\l]\}
   \approx \d_\l\G_L[\t]+\d_{\bar\l}\G_R[\bar\t]\, ,
$$
combine it with
\bea
&&\d_\l\G_L[\t]+\d_{\bar\l}\G_R[\bar\t]
   = -2\k\oint dx^\b\t^i\nabla_\b\l_i
    +2\bk\oint dx^\b\bar\t\pd_\b\bar\l\, ,                 \nn\\
&&\t^i\nabla_\b\l_i=\s^iA_{i\b}+\t^i\pd_\b\l_i\, ,         \nn
\eea
where  $\s^i=\ve^{ijk}\l_j\t_k$, and obtain
\bsubeq
\be
\{\tG[\t,\bar\t],\tG[\l,\bar\l]\}=\tG[\s,0]
  - 2\k\oint dx^\b\hat\t^i\pd_\b\hat\l_i
  + 2\bk\oint dx^\b\bar\t\pd_\b\bar\l\, .                  \lab{3.2a}
\ee
After introducing the Fourier modes
\bea
&&J_{im}:=\tG[\hat\t^i=e^{-im\vphi},\bt=0]
  \approx -2\k(\hA_{i2})_m\, ,                             \nn\\
&&K_m:=\tG[\hat\t^i=0,\bt=e^{-im\vphi}]
    \approx 2\bk(\bA_2)_m\, ,                              \nn
\eea
the PB algebra \eq{3.2a} takes the form of a Kac-Moody
extension of the $sl(2,R)\oplus u(1)$ Lie algebra:
\bea
&&i\{J^i_m,J^j_n\}=i\ve^{ij}{_k}J^k_{m+n}
                     +4\pi\k m\eta^{ij}\d_{m,-n}\, ,       \nn\\
&&i\{K_m,K_n\}=-4\pi\bk m\d_{m,-n}\, .
\eea
\esubeq

\subsection{Semidirect sum of Virasoro and {\boldmath$u(1)_{\rm KM}$}}

We now wish to examine another set of boundary conditions on $u^i$ and
$\bu$:
\bitem
\item[\bull] $u^i=-\th^i-\xi^\r A^i{_\r}$ and
       $\bu=-\xi^\r\bA_\r$ (at the boundary).
\eitem
These conditions are analogous to those used in the AdS sector of
Einstein's 3D gravity, but not identical \cite{4}; the presence of the
additional $\th^i$ term in $u^i$ will become clear soon. We begin the
analysis by discussing the symmetry structure of the $SL(2,R)$ sector.

\prg{{\boldmath $SL(2,R)$} sector.} Imposing the adopted gauge
conditions \eq{2.5}, the form of $\d\G_L[u]$ in \eq{2.7a} leads to
\be
\G_L= 2\k\int_0^{2\pi}d\vphi\left[(\hat\th+\xi^1 a_1)\hA_2
      +\frac{1}{2}\xi^2 (\hA_2)^2\right]\, ,               \lab{3.3}
\ee
where  $\th=:b^{-1}\hat\th b$.

To proceed, we impose two additional requirements:
\bsubeq\lab{3.4}
\be
\hA^1{}_2=0\,,\qquad \hA^-{}_2=-2C\,.                      \lab{3.4a}
\ee
They are of the same form as in the AdS sector of 3D gravity \cite{19},
but now, $C$ is an arbitrary constant. Using the residual gauge
symmetry \eq{2.6}, the invariance of these requirements implies
\bea
&&\hat\th^- +\xi^1 a_1^-=0\, ,                             \nn\\
&&\hat\th^1 +\xi^1 a_1^1=-\pd_2\xi^2\, ,                   \nn\\
&&C(\hat\th^+ +\xi^1 a_1^+)=\pd_2^2\xi^2\, .               \lab{3.4b}
\eea
\esubeq
The last equation shows why the additional $\th^i$ term is needed: with
the usual \grl\ choice $\th^i=0$ and for the standard ``gravitational"
value $a_1^+=0$ \cite{4}, we would have a too strong restriction
$\pd_2^2\xi^2=0$. Note that this is true even in \grl. As a consequence
of \eq{3.4b}, the integrand in \eq{3.3} is \emph{linearized}:
\bsubeq\lab{3.5}
\be
\G_L[\xi]=-\int d\vphi\xi^2\cM_L\, ,\qquad
\cM_L:=2\k C\hA^+{}_2\, .                                  \lab{3.5a}
\ee

The canonical algebra can be now derived using the transformation rule
\be
\d_\eta\cM_L=-2(\pd_2\eta^2)\cM_L
             -\eta^2\pd_2\cM_L-2\k\pd_2^3\eta^2\,.         \lab{3.5b}
\ee
\esubeq
Indeed, this rule implies
$$
\d_\eta\G_L[\xi]=\G_L[\s]+2\k\int d\vphi\xi^2\pd_2^3\eta^2\, ,
$$
with $\s^2=\eta^2\pd_2\xi^2-\xi^2\pd_2\eta^2$, and consequently,
\bsubeq
\be
\{\tG_L[\xi],\tG_L[\eta]\}
    =\tG_L[\s]+2\k\int d\vphi\xi^2\pd_2^3\eta^2\, .
\ee
Expressed in terms of the Fourier modes
$$
L'_m=\tG[\xi^2=e^{-im\vphi}]\approx -(\cM_L)_m\, ,
$$
the canonical algebra takes the form of a Virasoro algebra with
classical central charge:
\be
i\{L'_m,L'_n\}=(m-n)L'_{m+n}+4\pi \k m^3\d_{m,-n}\,.
\ee
\esubeq

\prg{The complete theory.} Going now to the $U(1)$ sector with
$\bu=-\xi^\r\bA_\r$ and imposing the additional restriction
$$
\bar a_1=0\, ,
$$
we obtain
\bea
&&\G_R=-2\bk\int_0^{2\pi}d\vphi\left[\xi^0\bar a_0\bA_2
  +\frac{1}{2}\xi^2(\bA_2)^2\right]\, ,                    \lab{3.7}\\
&&\d_\eta\bA_2=-\pd_2(\eta^2\bA_2)-\bar a_0\pd_2\eta^0\, . \nn
\eea
Combining \eq{3.7} with \eq{3.5}, we find that the complete surface
term has the form
\bsubeq\lab{3.8}
\bea
&&\G[\xi]:=\G_L[\xi]+\G_R[\xi]
     =-\int d\vphi\xi^0\cE-\int d\vphi\xi^2\cM\, ,         \nn\\
&&\cE=2\bk\bar a_0\bA_2\, ,\qquad
  \cM=\cM_L +\bk(\bA_2)^2\, .                              \lab{3.8a}
\eea
As before, the transformation rules
\bea
&&\d_\eta\cM=-2(\pd_2\eta^2)\cM-\eta^2\pd_2\cM
             -2\k\pd_2^3\eta^2-(\pd_2\eta^0)\cE\, ,        \nn\\
&&\d_\eta\cE=-(\pd_2\eta^2)\cE-\eta^2\pd_2\cE
               -2\bk\bar a_0^2\pd_2\eta^0\, ,
\eea
\esubeq
define the form of the complete PB algebra:
\be
\{\tG[\xi],\tG[\eta]\}=\tG[\s]+2\k\int d\vphi\xi^2\pd_2^3\eta^2
  +2\bk\bar a_0^2\int d\vphi\xi^0\pd_2\eta^0\, ,
\ee
where $\s^{\bar\a}=\eta^2\pd_2\xi^{\bar\a}-\xi^2\pd_2\eta^{\bar\a}$,
$\bar\a=0,2$. Expressed in terms of the the Fourier modes:
\bea
&&K_n=\tG[\xi^0=e^{-in\vphi},\xi^2=0]=-\cE_{n}\, ,         \nn\\
&&L_m=\tG[\xi^0=0,\xi^2=e^{-im\vphi}]=-\cM_{m}\, ,         \nn
\eea
the above PB algebra takes the form of the semidirect sum \vkm:
\bea
&&i\{L_m,L_n\}=(m-n)L_{m+n}+4\pi\k m^3\d_{m,-n}\, ,        \nn\\
&&i\{L_m,K_n\}=-nK_{m+n}\, ,                               \nn\\
&&i\{K_m,K_m\}=-4\pi\bk\bar a_0^2 m\d_{m,-n}\, .           \lab{3.10}
\eea

The gauge conditions \eq{2.5a} and \eq{2.5b} in conjunction with the
additional requirements \eq{3.4a} imply that the original set of $9+3$
gauge potentials $A^i{_\m}$ and $\bA_\m$ is reduced to just \emph{two
independent boundary degrees of freedom}, $\hA^+{}{_2}(\vphi)$ and
$\bA_2(\vphi)$. These are the only modes that appear in the CS surface
term \eq{3.8a}.

The basic content of our analysis is encoded in the form of the
\emph{surface term} \eq{3.8a} and the \emph{PB algebra of the
asymptotic generators} \eq{3.10}. These results will be compared to
those found in the asymptotic region of the spacelike stretched AdS
gravity.

\section{Spacelike stretched AdS gravity}
\setcounter{equation}{0} 

We now turn our attention to \tmgl, defined by the Lagrangian
\be
L_\tmg=2ab^i R_i-\frac{\L}{3}\,\ve_{ijk}b^ib^jb^k\,
  +a\m^{-1}L_\cs(\om)+\l^i T_i\, ,                         \lab{4.1}
\ee
where the notation is the same as in Ref. \cite{16}:  $\om^i$ is the
Lorentz connection and $b^i$ the orthonormal coframe, $R^i$ and $T^i$
are their associated field strengths, the curvature and torsion,
$L_\cs(\om)=\om^i d\om_i +\frac{1}{3}\ve_{ijk}\om^i\om^j\om^k$ is the
Chern-Simons Lagrangian for the connection, $\l^i$ is the Lagrange
multiplier that ensures the vanishing of torsion, and $a=1/16\pi G$. We
assume that $G$ is positive, while $\m$ remains arbitrary. By
construction, \tmgl\ is invariant under the local Poincar\'e
transformations.

The variation of the action with respect to $b^i,\om^i$ and $\l^i$,
yields the gravitational field equations:
\bsubeq\lab{4.2}
\bea
&&2aR_i-\L\ve_{ijk}b^jb^k+\nab\l_i=0\, ,                   \lab{4.2a}\\
&&2aT_i+2a\m^{-1}R_i+\ve_{imn}\l^mb^n=0\, ,                \lab{4.2b}\\
&&T_i=0\, ,                                                \lab{4.2c}
\eea
\esubeq
where $\nab\l_i=d\l_i+\ve_{ijk}\om^j\l^k$ is the covariant derivative
of $\l_i$.

In order to prepare a comparison between our CS theory and \tmgl, we
now give a brief account of the asymptotic structure of \tmgl\
and derive a new form of the surface terms.

\subsection{Spacelike stretched AdS asymptotics}

The spacelike stretched black hole is a solution of \tmgl\ (Appendix
B), which can be constructed as a discrete quotient space of the
spacelike stretched \ads3\ vacuum. This black hole generates an
interesting set of asymptotic states, the structure of which is
defined, in analogy with the AdS case, by the following requirements:
\bitem
\item[(a)] asymptotic configurations should include spacelike stretched
black hole geometries;\vsm
\item[(b)] they should be invariant under the action of
$SL(2,R)\times U(1)$, the isometry group of the spacelike stretched \ads3;\vsm
\item[(c)] asymptotic symmetries should have well-defined canonical
generators.
\eitem

In \cite{16}, this general approach was used to derive asymptotic
properties of the fields and gauge parameters. Based on the
requirements (a) and (b), the gravitational fields $b^i{_\m}$,
$\om^i{_\m}$ and $\l^i{_\m}$ are found to have the following asymptotic
form:
\bsubeq\lab{4.3}
\be
b^i{_\m}=\bar b^i{_\m}+B^i{_\m}\, ,\qquad
B^i{_\m}:=\left(
          \ba{ccc}
          \cO_1&\cO_3&\cO_1\\
          \cO_2&\cO_2&\cO_1\\
          \cO_1&\cO_3&\cO_0
          \ea
          \right)\, ,                                      \lab{4.3a}
\ee
\be
\om^i{_\m}=\bar\om^i{_\m}+\Om^i{_\m}\, ,\qquad
\Om^i{_\m}:=\left(
            \ba{ccc}
            \cO_1&\cO_3&\cO_0\\
            \cO_2&\cO_2&\cO_1\\
            \cO_1&\cO_3&\cO_0
            \ea
            \right)\, ,                                    \lab{4.3b}
\ee
\be
\l^i{_\m}= \bar\l^i{_\m}+\L^i{_\m}\, ,\qquad
\L^i{_\m}:=\left(
           \ba{ccc}
           \cO_1&\cO_3&\cO_0\\
           \cO_2&\cO_2&\cO_1\\
           \cO_1&\cO_3&\cO_0
           \ea
           \right)\, ,                                     \lab{4.3c}
\ee
\esubeq
where $\cO_n$ is a term that tends to zero as $r^{-n}$ or faster. The
expansion \eq{4.3} is an asymptotic expansion around the spacelike
stretched black hole vacuum $(\bar
b^i{_\m},\bar\om^i{_\m},\bar\l^i{_\m})$, displayed in Appendix B.

The subset of Poincar\'e gauge transformations that leave the
asymptotic configurations \eq{4.3} invariant defines the asymptotic
symmetry. The invariance of \eq{4.3} restricts the gauge parameters
$\xi^\m$ (translations) and $\th^i$ (Lorentz rotations) to have the
following form:
\bsubeq\lab{4.4}
\bea
&&\xi^0=\ell T(\vphi)-\frac{4\ell^2\n}{(\n^2+3)^2}\frac{1}{r}\pd_2^2 S
  +\cO_2\, ,\qquad \xi^1=-r\pd_2 S(\vphi)+\cO_0(\vphi)\, , \nn\\
&&\xi^2=S(\vphi)+\frac{2\ell^2}{(\n^2+3)^2}\frac{1}{r^2}\pd_2^2 S
        +\cO_3\, ,                                         \\
&&\th^0=-\frac{2\ell}{\sqrt{3(\n^2+3)(\n^2-1)}}
         \,\frac{1}{r}\,\pd_2^2S(\vphi)+\cO_2\, ,          \nn\\
&&\th^1=\frac{2\ell\sqrt{\n^2+3}}{3(\n^2-1)}
        \,\frac{1}{r}\,\pd_2 T(\vphi)+\cO_2\, ,            \nn\\
&&\th^2=-\frac{4\ell\n}{(\n^2+3)\sqrt{3(\n^2-1)}}
         \,\frac{1}{r}\,\pd_2^2S(\vphi)+\cO_2\, .
\eea
\esubeq
The leading order terms in \eq{4.4}, which are determined by just two
functions, $T(\vphi)$ and $S(\vphi)$, define the $(T,S)$
transformations; their time independence closely corresponds to the CS
boundary conditions \eq{2.2}. The sub-leading terms, those that remain
after imposing $T=S=0$, define the residual (or pure) gauge
transformations. The asymptotic symmetry is defined by the $(T,S)$
pair, ignoring all the residual gauge parameters.

In the expressions \eq{4.3} and \eq{4.4}, some typos appearing in
\cite{16} are corrected.

\subsection{Canonical PB algebra}

Using the adopted asymptotic conditions, the improved canonical
generator is given as \cite{16}:
\bea
&&\tG=G+\G\, ,                                             \nn\\
&&\G:=-\int_0^{2\pi}d\vphi\left(\ell T\cE^1+S\cM^1\right)\,,\lab{4.5}
\eea
where
\bsubeq\lab{4.6}
\bea
&&\cE^1=b^i{_0}\left[\frac{4a}{3}\om_{i2}+\l_{i2}
        -\frac{a}{3\ell\n}(2\n^2+3)b_{i2}\right]\, ,       \nn\\
&&\cM^1=b^i{_2}(2a\om_{i2}+\l_{i2})
        +\frac{a\ell}{3\n}\om^i{_2}\om_{i2}\, .            \lab{4.6a}
\eea
The improved generator $\tG$ is differentiable and has a finite value.
For $T=1$ and $S=1$, the surface term $\G$ defines the conserved
canonical charges, energy and angular momentum (for a background
independent approach to the conserved quantities, see \cite{20}).

The corresponding PB algebra is determined by the transformation laws
\bea
&&\d_0\cE^1=-S\pd_2\cE^1-(\pd_2 S)\cE^1
  -\frac{2a(\n^2+3)}{3\n}\pd_2 T\, ,                       \nn\\
&&\d_0\cM^1=-2(\pd_2 S)\cM^1-S\pd_2\cM^1-(\ell\pd_2 T)\cE^1
  -\frac{2a\ell(5\n^2+3)}{3\n(\n^2+3)}\pd_2^3S\, ,
\eea
\esubeq
Expressed in terms of the Fourier modes,
\bea
&&K_n:=\tG(T=e^{-in\vphi},S=0)\approx -\ell\cE^1_n\, ,     \nn\\
&&L_n:=\tG(T=0,S=e^{-in\vphi})\approx -\cM^1_n\, ,
\eea
the asymptotic canonical algebra of the spacelike stretched AdS gravity
reads:
\bsubeq\lab{4.8}
\bea
&&i\{L_m,L_n\}=(m-n)L_{m+n}+\frac{c_V}{12}m^3\d_{m,-n}\, , \nn\\
&&i\{L_m,K_n\}=-nK_{m+n}\, ,                               \nn\\
&&i\{K_m,K_n\}=-\frac{c_K}{12}m\d_{m,-n}\, ,               \lab{4.8a}
\eea
where
\be
c_V=\frac{(5\n^2+3)\ell}{G\n(\n^2+3)}\, ,\qquad
 c_K=\frac{(\n^2+3)\ell}{G\n}\, .                          \lab{4.8b}
\ee
\esubeq
One should note that this algebra is of the same form as the
corresponding CS algebra: \vkm.

\section{Boundary structure of the spacelike stretched AdS gravity}
\setcounter{equation}{0}

In this section, we introduce a set of specific asymptotic conditions
for the spacelike stretched AdS gravity, corresponding to the gauge
conditions introduced in the CS theory; then, we derive a new form of
the surface terms and discuss the boundary degrees of freedom.

\subsection{Specific asymptotic conditions}

In section 4.1, our intention was to construct the \emph{most general}
set of asymptotic conditions based on the requirements (a) and (b).
Here, we introduce  a set of the \emph{specific} (refined) asymptotic
conditions, \emph{compatible} with the general asymptotic structure.

We begin by noting that neither the black hole solution nor the leading
order asymptotic parameters depend on time. These properties can be
naturally extended by introducing the following refined asymptotic
conditions:
\be
(b^i{_\m},\om^i{_\m},\l^i{_\m})\quad{\rm ~and}\quad
(\xi^\m,\th^i)\quad {\rm are~time~independent}\, .         \lab{5.1}
\ee
In particular, \eq{5.1} implies that pure gauge parameters are time
independent.

Next, motivated again by the properties of the black hole solution
[to leading order, it is represented by the black hole vacuum
\eq{B.2}], we adopt the following conditions:
\be
\frac{\n}{\ell}b^i{_0}+\om^i{_0}=0\, ,\qquad
\frac{a}{\m\ell^2}(4\n^2-3)b^i{_0}-\l^i{_0}=0\,.           \lab{5.2}
\ee
One can verify that these conditions do not lead to any restriction on
the asymptotic para\-me\-ters. They can be considered as the gauge
conditions that are compatible with the spacelike stretched AdS
asymptotics, while canonically, the conditions \eq{5.2} are associated
to the first class constraints $(\pi_i{^0}{}',\Pi_i{^0})$ in \tmgl\
\cite{13}. Note the analogy between \eq{5.1}, \eq{5.2} and the CS
boundary conditions \eq{2.2} or gauge conditions \eq{2.5a}.

In the standard AdS gravity, the BTZ black hole satisfies the condition
$b^i{_+}/\ell+\om^i{_+}=0$, which is similar to the first condition in
\eq{5.2}. The difference stems from different asymptotic conditions in
the standard and spacelike stretched AdS gravity: using the CS
variables, these conditions can be expressed as $A_+=0$ and $A_0=0$,
respectively.

\subsection{A new form of the surface terms}

The surface terms $\cE^1$ and $\cM^1$ can be written in a form which
closely resembles the corresponding CS expressions \eq{3.8a}. Indeed,
the conditions \eq{5.1} and the equations of motion \eq{C.1a},
\eq{C.2a} and \eq{C.3a} imply
\bsubeq\lab{5.3}
\bea
&&\cE^1=-\frac{a[3(\n^2-1)]^{3/2}}{3\n\ell}\hB^2{_0}\,,    \\
&&\cM^1= \cM^+ +\frac{12\pi\ell^2}{c_K}(\cE^1)^2\, ,
\eea
where
\be
\cM^+:=-\frac{a\sqrt{3(\n^2-1)}}{2}
   \left(\frac{\hat\hB^2{_2}}{\n\ell}
         +\frac{2\sqrt{\n^2+3}}{3\ell}\hB^0{_2}
         +\frac{4}{3}\hat\hOm^2{_2}+\frac{1}{a}\hat\hL^2{_2}
         +\frac{2\sqrt{\n^2+3}}{3\n}\hat\hOm^0{_2}\right)\,.
\ee
\esubeq
Here, the first/second order sub-leading terms in the asymptotic
expansion of the fields are denoted by single/double hats. For
instance, the relation $B^2{_0}=\cO_1$ in \eq{4.3} is written as
$$
B^2{_0}=\frac{\hB^2{_0}}{r}+\frac{\hat\hB^2{_0}}{r^2}+\cdots\, ,
$$
and similarly for the other field components.

\subsection{Boundary degrees of freedom}

We are now going to prove the following statement:
\bitem
\item[$\bull$] in the spacelike stretched AdS sector of \tmgl, there
are two independent boundary degrees of freedom.
\eitem

Let us first note that the only sub-leading field modes that contribute
to the values of the conserved charges are those of the order $\cO_0$
and $\cO_1$. The connection $\om^i$ and the multiplier field $\l^i$ can
be expressed in terms of the triad by using the asymptotic expansion of
the equations of motion. The only second-order triad modes that appear
in these asymptotic relations are $\hhB^0{_0}$, $\hhB^2{_0}$,
$\hhB^2{_2}$ and $\hhB^1{_1}$ (see Appendix C). Thus, the number of
boundary modes is defined by the 9 first-order modes $\hB^i{_\m}$ plus
the additional 4 second-order modes.

However, not all of these modes are independent: there are 4
constraints \eq{C.1}, and 7 modes can be fixed by fixing the residual
gauge symmetry, defined by 7 residual gauge parameters in \eq{D.1}.
Thus, the number of independent boundary degrees of freedom is
$13-4-7=2$. They can be identified with $\cE^1$ and $\cM^+$, the
surface terms of the canonical generator, which are invariant under the
residual gauge transformations, see \eq{D.3}.

The \emph{boundary} degrees of freedom should not be confused with the
\emph{propagating} degrees of freedom. Thus, for instance, Einstein's
3D gravity is a topological theory without pro\-pa\-gating degrees of
freedom, but its AdS sector possesses two boundary degrees of freedom.

\section{Asymptotic relation between CS theory and {\boldmath\tmgl}}
\setcounter{equation}{0} 

We are now ready to establish a remarkable asymptotic relation between
the $SL(2,R)\times U(1)$ CS gauge theory and \tmgl:
\bitem
\item[$\bull$] asymptotic structures of the spacelike stretched AdS
gravity and the $SL(2,R)\times U(1)$ CS gauge theory can be identified
by adopting a natural asymptotic correspondence between their field
variables and coupling constants.
\eitem
The result holds when the gauge conditions \eq{2.5b} have the form
$a_1=T_1,\bar a_1=0$, and for a specific value of the constant $C$ in
\eq{3.4a}.

To prove the statement, we compare the \emph{asymptotic canonical
algebras} \eq{3.10} and \eq{4.8} and the corresponding \emph{surface
terms} \eq{3.8} and \eq{4.5} of the two theories, and find that these
structures coincide if we adopt the following \emph{asymptotic}
correspondence:
\bea
&&4\pi\bk\bar a_0^2\sim\frac{c_K}{12}\,,\qquad
  4\pi\k\sim\frac{c_V}{12}\, ,                             \nn\\
&&\cE\sim\cE^1\, ,\qquad  \cM\sim\cM^1\, .
\eea
Taking into account \eq{5.3}, the correspondence between the surface
terms reads
\be
2\bk\bar a_0\bA_2\sim \cE^1\, ,\qquad
2\k C\hA^+{}{_2}\sim \cM^+\, ,                             \lab{6.2}
\ee
or equivalently, when expressed in terms of the boundary modes,
\bea
&&\bA_2\sim
  -\frac{a[3(\n^2-1)]^{3/2}}{6\bk\bar a_0 \n\ell}\hB^2{_0}\,,\nn\\
&&\hA^+{}_2\sim -\frac{a\sqrt{3(\n^2-1)}}{4\k C}\left(
  \frac{\hhB^2{_2}}{\n\ell}+\frac{2\sqrt{\n^2+3}}{3\ell}\hB^0{_2}
  +\frac{4}{3}\hhOm^2{_2}+\frac{2\sqrt{\n^2+3}}{3\n}\hhOm^0{_2}
  +\frac{\hhL^2{_2}}{a}\right)\, .                         \nn
\eea

It is interesting to note that, under the adopted gauge and boundary
conditions, this correspondence can also be rewritten in a covariant
form (Appendix E):
\bsubeq\lab{6.3}
\bea
&&A^i{_\m}\sim\om^i{_\m}+\frac{3\n}{2(2\n+\sqrt{\n^2+3})}
  \left(\frac{3+2\n\sqrt{\n^2+3}}{3\ell\n}b^i{_\m}
        +\frac{1}{a}\l^i{_\m}\right)\, ,                   \lab{6.3a} \\
&&\bA_\m\sim\frac{\ell}{2\bk\bar a_0} b^i{_0}
  \left(\frac{4a}3\om_{i\m}+\l_{i\m}
        -\frac{a}{3\ell}\frac{2\n^2+3}{\n}b_{i\m}\right)\,.\lab{6.3b}
\eea
\esubeq
The transformation laws of $A^i{_\m}$ and $\bA^i{_\mu}$, induced by
\eq{6.3}, take the expected form:
\bsubeq
\bea
&&\d_0 A^i{_\m}\sim -\stackrel{A}{\nabla}_\m\th^i
  -(\pd_\mu\xi^\r)A^i{_\r}-\xi^\r\pd_\r A^i{_\m}\,,        \lab{6.4a}\\
&&\d_0 \bA^i{_\m}\sim -(\pd_\mu\xi^\r)\bA^i{_\r}
              -\xi^\r\pd_\r \bA^i{_\m}\,.                  \lab{6.4b}
\eea
\esubeq

To illustrate practical aspects of the established correspondence, we
note that in thermodynamic applications, one needs an action which is
both finite and differentiable \cite{21}. These properties are ensured
by the following procedure: first, if the value $I_{\rm bh}$ of the
action $I$ at the black hole configuration is divergent, we apply a
suitable \emph{regularization} to define $I_{\rm r}$, a finite piece of
$I$, and second, we construct the improved action $\tI=I_{\rm r}+B$,
where $B$ is a \emph{surface term} that ensures the differentiability
of $\tI$ under the adopted boundary conditions.

Let us now apply this procedure to the CS action \eq{2.1}. We begin by
noting that in the spacelike stretched AdS sector, the asymptotic
relations \eq{6.3} imply the gauge conditions \eq{2.5} and the
relations $\pd_0A_2=\pd_0\bA_2=0$. Then, one finds that $(I_\cs)_{\rm
bh}$ vanishes, $(I_\cs)_{\rm bh}\approx
(\k/3)\int\ve_{ijk}A^iA^jA^k\approx 0$, so that there is no need for
any regularization: $(I_\cs)_\reg =I_\cs$. After that, the improved
action $\tI_\cs$ is found to be of the form
$$
\tI_\cs=I_\cs +B_\cs\, ,\qquad
B_\cs:=-\bar\k\bar a_0\int_{\pd\cM}dt d\vphi\bA_2\, .
$$
In the Euclidean spacetime with the periodic time coordinate, $B_\cs$
is finite.

When the same procedure is applied to \tmgl, using \eq{4.1} and the
spacelike stretched boundary conditions, we find:
\bea
&&(I_\tmg)_\reg
  =I_\tmg+\int dt d\vphi\frac{a(\n^2+3)}{2\ell}r_\infty\, , \nn\\
&&\tI_\tmg=(I_\tmg)_\reg+B_\tmg\, ,\qquad
  B_\tmg=-\frac{1}{2}\int_{\pd\cM}dtd\vphi\cE^1\, ,         \nn
\eea
where $r_\infty$ is the value of $r$ at the boundary. Then, in view of
the asymptotic correspondence \eq{6.2}, the CS boundary term is seen to
coincide with its gravitational counterpart:
\be
B_\cs=B_\tmg\, ,                                            \lab{6.5}
\ee
and the on-shell values of the improved actions $\tI_\cs$ and
$\tI_\tmg$ are identical. This result gives a deeper insight into the
correspondence of the two theories, extending it from an asymptotic
relation between fields and coupling constants, to the level of
equality of the boundary terms needed to improve the regularized
actions. The equality \eq{6.5} might lead to a simplified approach to
the gravitational entropy, which is, on the other hand, closely related
to the question of warped AdS/CFT correspondence in \tmgl\
\cite{10,16,17}.

\section{Concluding remarks}

In this paper, we compared the asymptotic structures of the spacelike
stretched AdS gravity and the $SL(2,R)\times U(1)$ CS gauge theory.

(1) We studied the asymptotic properties of the $SL(2,R)\times U(1)$ CS
gauge theory in the canonical formalism. By imposing a suitable set of
the gauge and boundary conditions, we calculated two conserved charges
and found that the boundary symmetry is described by the \vkm\ PB
algebra with central charges.

(2) This result shows a remarkable resemblance with the properties of
the spacelike stretched AdS gravity. Indeed, by comparing the boundary
canonical algebras and the surface terms of the improved generators in
the two theories, one finds that their asymptotic structures can be
identified by adopting a natural mapping between the respective
coupling constants and field variables.
Thus, in spite of the fact that \tmgl\ is not a topological theory, the
asymptotic structure of its \emph{spacelike stretched AdS sector} can
be faithfully represented by the $SL(2,R)\times U(1)$ CS gauge theory.
Note that this result holds \emph{only asymptotically}, not in the
bulk. It represents a natural extension of the known asymptotic
correspondence between the \emph{AdS sector} of \tmgl\ and another
topological gauge theory---the Mielke-Baekler 3D gravity with vanishing
torsion \cite{13}, or equivalently, the $SL(2,R)\times SL(2,R)$
CS gauge theory \cite{22}.

(3) As indicated by equality of the boundary terms needed to improve
the regularized CS and \tmgl\ Euclidean actions, the asymptotic CS
representation of \tmgl\ might be a useful tool in clarifying the
status of the hypothesis conjectured by Anninos et al \cite{10}.

\section*{Acknowledgements} 

This work was partially supported by the Serbian Science Foundation
under Grant No. 141036.

\appendix

\section{{\boldmath The $sl(2,R)$} Lie algebra: conventions}
\setcounter{equation}{0} 

For the basis of the fundamental matrix representation of the $sl(2,R)$
Lie algebra (real, traceles, $2\times 2$ matrices), we choose:
$$
T_0=\frac{1}{2}\left(\ba{cc}
                       0 & 1 \\
                      -1 & 0
                     \ea\right)\, ,\qquad
T_1=\frac{1}{2}\left(\ba{cc}
                       1 & 0 \\
                       0 &-1
                     \ea\right)\, ,\qquad
T_2=\frac{1}{2}\left(\ba{cc}
                       0 & 1 \\
                       1 & 0
                     \ea\right)\, .
$$
In this basis, the components of the Cartan metric are
$\eta_{ij}=-2{\rm Tr}(T_iT_j)=(+,-,-)$, and the form of the Lie algebra
is $[T_i,T_j]=\ve_{ij}{^k}T_k$, with $\ve_{012}=+1$. In these
conventions, the gauge potential can be represented as
$$
A=A^iT_i=\frac{1}{2}\left(\ba{cc}
                             A^1  & A^+  \\
                             -A^- & -A^1
                          \ea\right)\, ,
$$
where $A^\pm=A^0\pm A^2$ are the light-cone components of $A^i$.

\section{The spacelike stretched black hole}
\setcounter{equation}{0} 

The spacelike stretched black hole is a solution of \tmgl, which
represents a discrete quotient of the spacelike stretched \ads3\
vacuum. Using the notation $\L=-a/\ell^2$, $\n=\m\ell/3$, the metric of
the black hole in Schwarzschild-like coordinates is given by:
\be
ds^2=N^2dt^2-B^{-2}dr^2-K^2(d\vphi+N_\vphi dt)^2\, ,       \lab{B.1}
\ee
where \bea &&N^2=\frac{(\n^2+3)(r-r_+)(r-r_-)}{4K^2}\, ,\qquad
  B^2=\frac{4N^2K^2}{\ell^2}\, ,                           \nn\\
&&K^2=\frac{r}{4}\left[3(\n^2-1)r+(\n^2+3)(r_++r_-)
                 -4\n\sqrt{r_+r_-(\n^2+3)}\right]\, ,      \nn\\
&&N_\vphi=\frac{2\n r-\sqrt{r_+r_-(\n^2+3)}}{2K^2}\, .     \nn
\eea
The metric is defined for $\n^2>1$.

As shown in \cite{16}, we can use \eq{B.1} to calculate the simple
diagonal form of the triad field $b^i$, then, the connection $\om^i$ is
determined by the condition of vanishing torsion, and finally, the
solution for the multiplier $\l^i$ is found from \eq{4.2b}. The triple
$(b^i,\om^i,\l^i)$ defined in this way represents the spacelike
stretched black hole in the first-order formalism. The corresponding
black hole vacuum $(\bar b^i,\bar\om^i,\bar\l^i)$ is defined by the
conditions $r_+=r_-=0$:
\bsubeq\lab{B.2}
\bea
&&\bar{b}^i{_\m}=\left(
\ba{ccc}
\sqrt{\dis\frac{\n^2+3}{3(\n^2-1)}}&0&0 \\
0&\dis\frac{1}{\sqrt{\n^2+3}}\frac{\ell}{r}&0 \\
\dis\frac{2\n}{\sqrt{3(\n^2-1)}}&0&\dis\frac{\sqrt{3(\n^2-1)}}2r
\ea
\right),
\eea
\bea
&&\bar\om^i{_\m}=\left(
\ba{ccc}
-\dis\frac{\n}{\ell}\sqrt{\frac{\n^2+3}{3(\n^2-1)}}&0&
          -\dis\sqrt{3(\n^2+3)(\n^2-1)}\frac{r}{2\ell} \\
0&\dis\frac{\n}{\sqrt{\n^2+3}}\frac{1}r&0 \\
-\dis\frac{2\n^2}{\ell\sqrt{3(\n^2-1)}}&0&
          -\dis{\n}\sqrt{3(\n^2-1)}\frac{r}{2\ell}
\ea
\right),
\eea
\bea
&&\bar\l^i{_\m}= \frac{2a}{\m}\left(
\ba{ccc}
\dis\frac{4\n^2-3}{2\ell^2}\sqrt{\frac{\n^2+3}{3(\n^2-1)}}&0&
            \dis\frac{\n}{\ell^2}\sqrt{3(\n^2+3)(\n^2-1)}\,r \\
0&\dis\frac{3-2\n^2}{2\ell\sqrt{\n^2+3}}\frac 1{r}&0\\
\dis\frac{(4\n^2-3)\n}{\ell^2\sqrt{3(\n^2-1)}}&0&
            \dis\frac{3(2\n^2+1)}{4\ell^2}\sqrt{3(\n^2-1)}\,r
\ea
\right).\qquad
\eea
\esubeq

\section{Asymptotic expansion of the equations of motion}
\setcounter{equation}{0} 

Let us now explore the equations of motion in the asymptotic region.

We start with equation \eq{4.2c} which, in conjunction with the
specific asymptotic conditions \eq{5.1} and \eq{5.2}, leads to four
constraints on the triad modes. Two of them involve the first-order
modes,
\bsubeq\lab{C.1}
\bea
&&\hB^0{_0}-\frac{2\n}{\sqrt{\n^2+3}}\hB^2{_0}=0\,,\nn\\
&&\hB^1{_1}+\frac{2\n\ell\sqrt{3(\n^2-1)}}{(\n^2+3)^{3/2}}\hB^2{_0}
  +\frac{2\ell}{\sqrt{3(\n^2-1)(\n^2+3)}}\hB^2{_2}=0\,,    \lab{C.1a}
\eea
and the remaining two contain second-order modes:
\bea
&&\hhB^0{_0}-\frac{2\n}{\sqrt{\n^2+3}}\hhB^2{_0}+\frac{1}{2}
  \left[\frac{3(\n^2-1)}{\n^2+3}\right]^{3/2}(\hB^2{_0})^2=0\,,\nn\\
&&\hhB^1{_1}+\frac{\ell\sqrt{3(\n^2-1)}}{\n^2+3}\hhB^0{_0}
  +\frac{2\ell}{\sqrt{3(\n^2+3)(\n^2-1)}}\hhB^2{_2}
  -\frac{4\ell\n}{(\n^2+3)\sqrt{3(\n^2-1)}}\hB^0{_2}       \nn\\
&&-\frac{\sqrt{\n^2+3}}\ell(\hB^1{_1})^2
  +\frac{2\ell}{\n^2+3}\hB^0{_0}\hB^2{_2}=0\,.             \lab{C.1b}
\eea
\esubeq
The remaining equations are algebraic relations containing the
connection modes. For the first-order modes, one obtains:
\bsubeq\lab{C.2}
\bea\lab{C.2a}
&&\hOm^0{_2}+\frac{3(\n^2-1)}{2\ell}\hB^0{_0}
  +\frac{\sqrt{\n^2+3}}{\ell}\hB^2{_2}=0\,,                \nn\\
&&\hOm^2{_2}+\frac{3(\n^2-1)}{2\ell}\hB^2{_0}
  +\frac{\n}{\ell}\hB^2{_2}=0\, ,                          \nn\\
&&\hOm^1{_1}-\frac{\n}{\ell}\hB^1{_1}
  -\sqrt{\frac{3(\n^2-1)}{\n^2+3}}\hB^2{_0}=0\,,           \nn\\
&&2\hB^1{_0}+\frac{2\n}{\sqrt{3(\n^2-1)}}\left(\frac\n\ell\hB^0{_1}
  -\hOm^0{_1}\right)-\sqrt{\frac{\n^2+3}{3(\n^2-1)}}
   \left(\frac\n\ell\hB^2{_1}-\hOm^2{_1}\right)=0\,,       \nn\\
&&\pd_2\hB^2{_0}-\sqrt{\frac{\n^2+3}{3(\n^2-1)}}
  \left(\frac\n\ell\hB^1{_2}-\hOm^1{_2}\right)
  +\frac{\sqrt{3(\n^2+3)(\n^2-1)}}{2\ell}\hB^1{_0}=0\,,    \nn\\
&&\pd_2\hB^1{_1}+\hB^1{_2}
  -\frac{\sqrt{3(\n^2+3)(\n^2-1)}}{2\ell}\hB^2{_1}
  +\frac{\sqrt{3(\n^2-1)}}
           2\left(\frac\n\ell \hB^0{_1}-\hOm^0{_1}\right)=0\,,
\eea
and the equations containing the second-order modes are:
\bea
&&2\hhB^0{_0}+\frac{2\n}{\sqrt{3(\n^2-1)}}\left(\frac\n\ell\hhB^1{_1}
  -\hhOm^1{_1}\right)+\hB^2{_0}\left(\frac\n\ell\hB^1{_1}
  -\hOm^1{_1}\right)=0\,,                                  \nn\\
&&\frac{\sqrt{3(\n^2+3)(\n^2-1)}}{2\ell}\hhB^1{_1}+\hhB^2{_2}
  -\frac{\ell}{\sqrt{\n^2+3}}
          \left(\frac{\n}{\ell}\hB^0{_2}+\hhOm^0{_2}\right)
  -\hB^1{_1}\hOm^0{_2}=0\, ,                               \nn\\
&&-\frac{\sqrt{3(\n^2-1)}}2\left(\frac\n\ell\hhB^1{_1}
  -\hhOm^1{_1}\right)+\frac\ell{\sqrt{\n^2+3}}
  \left(\frac\n\ell\hhB^2{_2}+\hhOm^2{_2}\right)           \nn\\
&&-\hB^0{_2}+\hOm^1{_1}\hB^2{_2}+\hOm^2{_2}\hB^1{_1}=0\,.  \lab{C.2b}
\eea
\esubeq

Similarly, starting from equation \eq{4.2b}, we find the following
independent algebraic relations for the the first-order multiplier
modes:
\bsubeq\lab{C.3}
\bea
&&\hL^0{_2}-\frac{2a}{3\n\ell}\left[
  \frac{3(\n^2-1)(5\n^2+3)}{4\n}\hB^0{_0}
  +2\n\sqrt{\n^2+3}\hB^2{_2}\right]=0\,,                   \nn\\
&&\hL^2{_2}-\frac{2a}{3\n\ell}\left[
  3(\n^2-1)\sqrt{\n^2+3}\hB^0{_0}
  +\frac{3(2\n^2+1)}2\hB^2{_2}\right]=0\, ,                \nn\\
&&\hL^1{_1}+\frac{a}{3\n\ell}(2\n^2-3)\hB^1{_1}=0\,.       \nn\\
&&-\frac{2(\n^2-1)}\ell\hB^1{_0}
  +\frac{(2\n^2-3)}{3\n\ell}\hB^1{_2}+\frac 1a\hL^1{_2}=0\,,\nn\\
&&\frac{2\n^2-3}{3\n\ell}\left(\sqrt{\n^2+3}\hB^2{_1}
  -2\n\hB^0{_1}\right)+\frac{\sqrt{\n^2+3}}{a}\hL^2{_1}
  -\frac{2\n}{a}\hL^0{_1}=0\,,                             \nn\\
&&-\frac{\sqrt{3(\n^2-1)}}{6}
  \left[\frac{3(2\n^2+1)}{2\ell\n}\hB^0{_1}-\hOm^0{_1}
  -\frac{\sqrt{\n^2+3}}{\n}\left(\frac{2\n}\ell \hB^2{_1}
  -\hOm^2{_1}\right)\right]                                \nn\\
&&+\frac\ell{3\n}\left(\hOm^1{_1}+\hOm^1{_2}\right)
  -\frac{\sqrt{3(\n^2-1)}}{4a}\hL^0{_1}=0\,,               \lab{C.3a}
\eea
and similarly for the second-order modes:
\bea
&&\hhL^1{_1}+\frac{a(2\n^2-3)}{3\n\ell}\hhB^1{_1}=0\,,\nn\\
&&\frac{(2\n^2+1)\sqrt{3(\n^2-1)}}{4\n\ell}\hhB^1{_1}
  +\frac{(3-2\n^2)}{6\n\sqrt{\n^2+3}}\hhB^2{_2}
  -\frac{\sqrt{3(\n^2-1)}}{6}\hhOm^1{_1}                   \nn\\
&&-\frac{\ell}{3\n}\hhOm^0{_2}
  +\frac{\ell}{3\sqrt{\n^2+3}}\hhOm^2{_2}
  +\frac{\sqrt{3(\n^2-1)}}{4a}\hhL^1{_1}
  +\frac{\ell}{2a\sqrt{\n^2+3}}\hhL^2{_2}                  \nn\\
&&+\frac{\ell}{3\n}\hOm^1{_1}\hOm^2{_2}
  +\frac{1}{2a}\hB^2{_2}\hL^1{_1}+\frac{1}{2a}\hB^1{_1}\hL^2{_2}=0\,,\nn\\
&&-\frac{2\sqrt{3(\n^2+3)(\n^2-1)}}{3\ell}\hhB^1{_1}
  +\frac{\ell}{3\n}\hhOm^2{_2}
  -\frac{\ell}{a\sqrt{\n^2+3}}\hhL^0{_2}
  +\frac{2\n^2-3}{3\n\sqrt{\n^2+3}}\hB^0{_2}               \nn\\
&&-\frac{\sqrt{3(\n^2-1)}}{6}
   \left(\hOm^0{_1} -\frac{\sqrt{\n^2+3}}{\n}\hOm^2{_1}\right)
  -\frac{1}{a}\hB^1{_1}\hL^0{_2}=0\,.                     \lab{C.3b}
\eea
\esubeq

The remaining equations \eq{4.2a} do not lead to any new relations.
Thus, we see that $\hOm^i{_\m},\hhOm^i{_\mu}$ and
$\hL^i{_\m},\hhL^i{_\m}$ can be expressed in terms of $\hB^i{_\m}$,
$\hhB^i{_\mu}$.

\section{Residual gauge transformations}
\setcounter{equation}{0} 

In this appendix, we calculate the action of the residual gauge
transformations on the triad modes. These transformations are defined
by \eq{4.4} with $T=S=0$ and are denoted by $\hd_0$. For the
first-order triad modes we have:
\bsubeq\lab{D.1}
\bea
&&\hd_0 \hB^0{_0}\equiv \hd_0\hB^2{_0}=0\,,                \nn\\
&&\hd_0 \hB^0{_1}=-\frac{\ell}{\sqrt{\n^2+3}}\hth^2
  +2\sqrt{\frac{\n^2+3}{3(\n^2-1)}}\hxi^0\,,               \nn\\
&&\hd_0\hB^0{_2}=\frac{\sqrt{3(\n^2-1)}}2\hth^1\,,         \nn\\
&&\hd_0\hB^1{_0}=\frac{2\n}{\sqrt{3(\n^2-1)}}\hth^0
  -\sqrt{\frac{\n^2+3}{3(\n^2-1)}}\hth^2\,,                \nn\\
&&\hd_0\hB^1{_1}=\frac{\ell}{\sqrt{\n^2+3}}\hxi^1\,,       \nn\\
&&\hd_0\hB^1{_2}=\frac{\sqrt{3(\n^2-1)}}2\hth^0
  -\frac{\ell}{\sqrt{\n^2+3}}\pd_2\hxi^1\,,                \nn\\
&&\hd_0\hB^2{_1}=-\frac{\ell}{\sqrt{\n^2+3}}\hth^0
  +\frac{4\n}{\sqrt{3(\n^2-1)}}\hxi^0
  +\frac 32\sqrt{3(\n^2-1)}\hxi^2\,,                       \nn\\
&&\hd_0\hB^2{_2}=-\frac{\sqrt{3(\n^2-1)}}2\hxi^1\,.
\eea
Hence, there is only one component $\hB^2{_0}$ [or equivalently
$\hB^0{_0}$, because of \eq{C.1a}] which remains invariant under
the residual gauge transformations.

For the second-order triad modes (with non-zero vacuum values), the
transformation laws read:
\bea
&&\hd_0\hhB^0{_0}=\frac{2\n}{\sqrt{3(\n^2-1)}}\hth^1
                  +\hxi^1\hB^0{_0}\, ,                     \nn\\
&&\hd_0\hhB^2{_0}=\sqrt{\frac{\n^2+3}{3(\n^2-1)}}\hth^1
                  +\xi^1\hB^2{_0}\, ,                      \nn\\
&&\hd_0\hhB^1{_1}=\frac{2\ell}{\sqrt {\n^2+3}}\hhxi^1
                  +2\hxi^1\hB^1{_1}\, ,                    \nn\\
&&\hd_0\hhB^2{_2}=-\frac{\sqrt{3(\n^2-1)}}2\hhxi^1\, .
\eea
\esubeq

In a similar way, we can find the residual gauge transformations for
the connection and the multiplier modes. The modes that appear in the
expressions for the asymptotic charges \eq{5.3} transform in the
following manner:
\bea
&&\hd_0\hhOm^2{_2}=-\frac{\sqrt{3(\n^2-1)(\n^2+3)}}{2\ell}\hth^1
  +\frac{\n\sqrt{3(\n^2-1)}}{2\ell}\hhxi^1\, ,             \nn\\
&&\hd_0\hhOm^0{_2}=-\frac{\n\sqrt{3(\n^2-1)}}{2\ell}\hth^1
  +\frac{\sqrt{3(\n^2+3)(\n^2-1)}}{2\ell}\hhxi^1\,,        \nn\\
&&\hd_0\hhL^2{_2}=\frac{2a\sqrt{3(\n^2-1)(\n^2+3)}}{3\ell}\hth^1
  -\frac{a(2\n^2+1)\sqrt{3(\n^2-1)}}{2\ell\n}\hhxi^1\, .   \lab{D.2}
\eea
Using these results, one can verify that the asymptotic charges are
invariant under the residual gauge transformations:
\be
\hd_0\cE^1=0\,,\qquad \hd_0\cM^1=0\, .                     \lab{D.3}
\ee
Indeed, the invariance of $\cE^1$ follows from $\hd_0\hB^2{_0}=0$, while
the transformation laws for $\hB^0{_2}$ and $\hhB^2{_2}$ in \eq{D.1},
together with the relations \eq{D.2}, imply $\hd_0\cM^+=0$, hence
$\hd_0\cM^1=0$.

\section{Derivation of the asymptotic relations \eq{6.3}}
\setcounter{equation}{0} 

In this appendix, we derive the asymptotic formulas \eq{6.3}, relating
the field variables of the CS theory to those of the spacelike
stretched AdS gravity, in the asymptotic region.

\subsection*{{\boldmath $SL(2,R)$} sector}

Let us first consider the $SL(2,R)$ sector of the theory. Radial
coordinates in the CS theory and \tmgl, $\r$ and $r$ respectively, are
not identical. They are connected by $\ell e^{\r/\ell}\sim r$. Hence,
for $a_1=T_1$, we have
$$
b\equiv e^{\r T_1/\ell}=\left(\ba{cc}
                  \sqrt{r/\ell} & 0 \\
                  0 & \sqrt{\ell/r}
                  \ea\right)\, .
$$
Using the gravitational radial coordinate $r$ also in the CS theory, we
find
\be
A^i{_r}=\frac{\d^i_1}{r}\, , \qquad
\hA^+{_2}=\frac{r}{\ell} A^+{_2}\, ,\qquad
\hA^-{_2}=\frac{\ell}{r} A^-{_2}\, ,
\ee
and consequently:
$$
A^+{_2}\sim-\frac{a\ell\sqrt{3(\n^2-1)}}{4\k C}
  \left(\frac{b^2{_2}}{\n\ell}+\frac{2\sqrt{\n^2+3}}{3\ell}b^0{_2}
  +\frac{4}{3}\om^2{_2}+\frac{2\sqrt{\n^2+3}}{3\n}\om^0{_2}
  +\frac{\l^2{_2}}{a}\right)\, .
$$
Next, we note that \eq{C.1b}, \eq{C.2b} and \eq{C.3b} imply:
$$
A^+{_2}\sim -\frac{a\ell\sqrt{3(\n^2-1)}}{4\k C}
  \left[\left(\frac{1}{\n\ell}+\frac{2\sqrt{\n^2+3}}{3\ell}\right)b^+{_2}
  +\left(\frac{4}{3}+\frac{2\sqrt{\n^2+3}}{3\n}\right)\om^+{_2}
  +\frac{\l^+{_2}}{a}\right]\, .
$$
This result motivates us to assume  the following general
correspondence:
\be
A^i{_\mu}\sim-\frac{a\ell\sqrt{3(\n^2-1)}}{4\k C}\left[
  \left(\frac{1}{\n}+\frac{2\sqrt{\n^2+3}}3\right)\frac{b^i{_\m}}{\ell}
  +\left(\frac{4}{3}+\frac{2\sqrt{\n^2+3}}{3\n}\right)\om^i{_\m}
  +\frac{\l^i{_\m}}{a}\right]\, ,                          \lab{E.2}
\ee
where $4\k=c_V/12\pi$. To prove this assumption, we examine its
validity for all values of the indices, using the adopted gauge and
asymptotic conditions.

For $\mu=0$, $A^i{_0}$ vanishes as a consequence of \eq{2.5a}, while
\eq{5.2} implies that the rhs of \eq{E.2} also vanishes.

For $\m=2$ and $i=1$, the first additional requirement in \eq{3.4a}
yields $A^1{_2}=\hA^1{_2}=0$, while \eq{4.3} implies that the
rhs of \eq{E.2} $\sim \cO_2$. For $\m=2$ and $i=-$, we use \eq{4.3}
and the second requirement in \eq{3.4a}, which imply that \eq{E.2} is
satisfied for
$$
C=\mp\frac{(\n^2+3)\sqrt{3(\n^2-1)}}{2(2\n+\sqrt{\n^2+3})}\,.
$$
Choosing the negative value of $C$, \eq{E.2} takes the form \eq{6.3a}.

Finally, for $\m=1$, the conditions \eq{4.3} imply that \eq{6.3a} is
identically satisfied.

The transformation law \eq{6.4a} of $A^i{_\m}$, induced by the relation
\eq{6.3a}, would not have been correct if we had chosen the plus sign
for $C$.

\subsection*{{\boldmath $U(1)$} sector}

By using \eq{4.6a}, the asymptotic expression for $\bA_2$ can be
written as:
$$
\bA_2\sim\frac{\ell}{2\bk\bar a} b^i{_0}
  \left(\frac{4a}{3}\om_{i2}+\l_{i2}
        -\frac{a}{3\ell}\frac{2\n^2+3}{\n}b_{i2}\right)\, .
$$
This relation can be consistently generalized to \eq{6.3b}. Indeed, for
$\m=1$, the rhs of \eq{6.3b} $\sim \cO_2$ as a consequence of \eq{4.3},
in agreement with the gauge condition $\bA_1=0$. Similarly, using the
refined asymptotic conditions \eq{5.2}, we find the expected result:
$$
\bA_0\sim -\frac{a\ell(\n^2+3)}{3\n\k\bar a_0}g_{00}\equiv\bar a_0\,.
$$

\end{document}